# Study of the electronic structure of short chain oligothiophenes

C. Grazioli[a,b], O. Baseggio[a], M. Stener[a], G. Fronzoni*[,a]

M. de Simone[c]

M. Coreno[b]

A Guarnaccio[d], S. Orlando[d], A. Santagata[d], M. D'Auria[e]

[a] Department of Chemical and Pharmaceutical Sciences, University of Trieste, 34127 Trieste, Italy

[b] CNR-ISM, Sincrotrone Trieste, Basovizza, I-34149 Trieste, Italy

[c] CNR-IOM, Laboratorio TASC, Sincrotrone Trieste, Basovizza, I-34149 Trieste, Italy

[d] CNR-ISM, Tito Scalo, C/da S. Loja, 85050 Tito Scalo, Potenza, Italy

[e] Science Department, University of Basilicata, 85100 Potenza, Italy

**Abstract**

The electronic structure of short-chain thiophenes (thiophene, 2,2'-bithiophene and 2,2':5',2''-terthiophene) in the gas phase has been investigated by combining the outcomes of Near-Edge X-ray-Absorption Fine-Structure (NEXAFS) and X-ray Photoemission Spectroscopy (XPS) at the C K-edge with those of density functional theory (DFT) calculations. The calculated NEXAFS spectra provide a comprehensive description of the main experimental features and allow their attribution. The evolution of the C1s NEXAFS spectral features is analyzed as a function of the number of thiophene rings; a tendency to a stabilization for increasing chain length is found. The computation of the binding energy allows to assign the experimental XPS peaks to the different carbon sites on the basis of both the inductive effects generated by the presence of the S atom as well as of the differential aromaticity effects.

Corresponding Authors:

* E-mail : fronzoni@units.it

**Introduction**

There has been a long-standing interest in the development of opto-electronic devices based on semiconducting or $\pi$-conjugated materials, as alternative to inorganic semiconductors, for advantages that include low cost manufacture, simple processing, mechanical flexiblility.[1,2,3] In the field of organic electronic devices the thiophene based polymers and oligomers have been particularly investigated and frequently used for their unique electronic, optical and redox properties.[4,5,6] Usually the thiophene rings are connected via the 'α-carbon' atoms, adjacent to the sulphur atom (α-conjugated polymer and oligomer). The high polarizability of the S atom in the rings stabilizes the conjugated chain conferring excellent charge transport properties which are of fundamental importance for applications in organic and molecular electronics.[7,8] However, the rapid developments towards applications left behind the basic understanding of many of the physical processes involved and hence detailed studies of model systems appear to be a special need. In this respect, the oligothiophenes represent excellent model compounds to provide insights into the properties of the more complex corresponding polymers and the monitoring of their properties in dependence on the increasing number of thiophenes can allow to establish valuable relationships and extrapolations to the polymer [9,10]. In particular, an extensive comprehension of the electronic structure of short oligomeric systems is essential to correctly interpret and exploit the mechanisms which underlie their potential device applications, however, to date, a systematic study of the electronic structure in short chain thiophenes - starting from the most simple building blocks and increasing the complexity by adding more components - is still missing.

Spectroscopies involving core electron transitions are useful tools to investigate the electronic structure of single free molecules and even of very complex systems: the strongly localized nature of the core hole on a specific site makes the spectral features very sensitive to the local environment of the absorbing atom in a very narrow spatial range, which depends on the extent of the overlap between the core initial state and the atomic valence component of the excited atom in the LCAO final wavefunction.[11] The Near Edge X-ray Absorption Spectroscopy (NEXAFS) is an ideal method to probe the electronic properties of all kinds of materials[11] through the fine structure in the spectra which corresponds to the transitions from the core orbitals to the unoccupied orbitals. The excitation process is governed by dipole selection rules and the oscillator strengths (transition intensities) are directly connected with the atomic site component of the virtual orbitals which is dipole allowed. However the interpretation of the NEXAFS spectral features is not straightforward and the support of theoretical calculations becomes very important to determine the nature of the electronic transitions and to assign the spectral features on a solid and reliable ground. Closely related to NEXAFS is X-ray Photoelectron Spectroscopy (XPS), which measures the ionization energies (IEs) of the core electrons. Also IEs are very sensitive to the local chemical and physical environment of the ionized atomic site and depend on two main contributions, the electron density (initial state effect) and the relaxation (final state effect) following the core hole formation, which can be reproduced by DFT calculations.[12] At variance with NEXAFS, in XPS it is difficult to calculate the intensity of the primary lines since the unbound photoelectron wavefunction of the final ionic state obeys boundary conditions which are not supported by conventional basis sets employed in quantum chemistry, like Gaussian or Slater functions. Therefore, if the photoelectron energy is not too low, the relative intensity among different primary lines can be simply assumed to be proportional to the number of equivalent atoms of the same type.

In this paper, we report an investigation of the electronic properties of thiophene (1T), 2,2'-bithiophene (2T) and 2,2':5',2''-terthiophene (3T) (shown in Fig.1), which represent the monomer, dimer and trimer of polythiophene respectively, through a combined experimental and theoretical analysis of the C1s NEXAFS and XPS spectra in the gas phase. These systems are challenging because they include at the same time strong electronic effects related to the aromatic stabilization as well as inductive effects generated by the presence of the S atom. The measured spectra have been rationalized with DFT calculations which are suitable to treat systems of increasing size at the same level of accuracy, allowing to follow the evolution of the spectral features with the increasing number of thiophene units in the chain. We remake that an analogous experimental and theoretical analysis of the S L-edge would be in order to complete the electronic characterization of the thiophenes. However when degenerate core holes are considered, like 2p orbitals, the coupling between different excitation channels requires theoretical approaches beyond the single determinant level presently employed [13]. Furthermore also the Spin Orbit (SO) coupling has to be included in the calculations in order to describe both the L2 and L3 series of transitions, which are expected to be strongly overlapped in S2p compounds due to the small SO energy splitting [14,15], rendering the spectral attribution quite demanding. The complexity of this issue therefore deserves a future separate work.

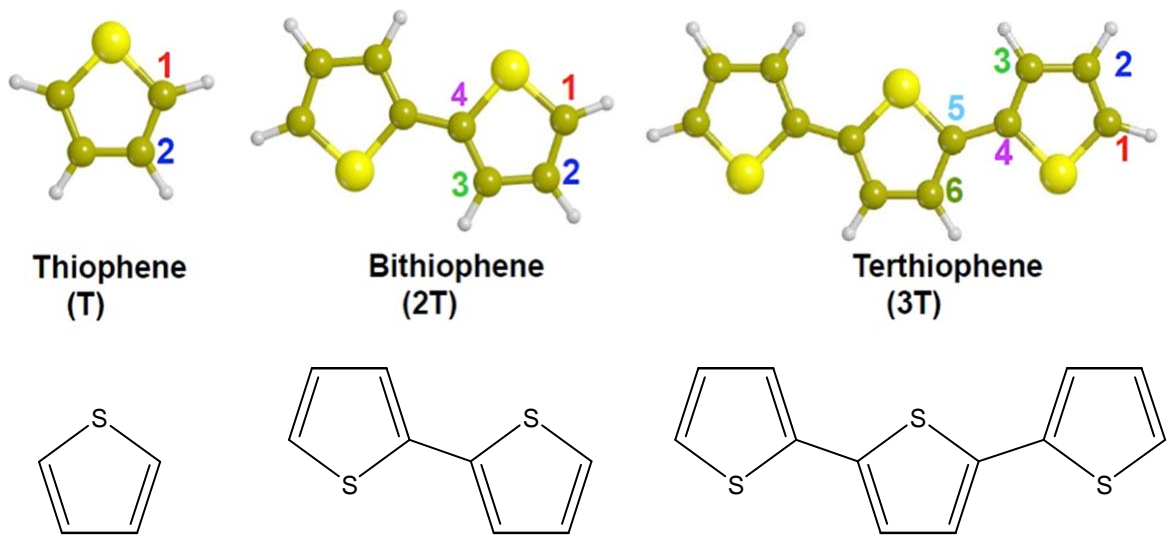

*Figure 1. Upper panel : Schematic illustration of thiophene (1T), 2,2'-bithiophene (2T) and 2,2':5',2''-terthiophene (3T). The nonequivalent carbon atoms are labelled. Lower panel: Structures with the evidence of the double bonds. The molecular plane corresponds to the xy plane*

**Experimental section**

The measurements on the gas-phase samples were performed at the Gas Phase beamline of the Elettra synchrotron in Trieste[16] using a Scienta SES-200 electron analyzer[17] mounted at the magic angle with respect to the electric vector of the linearly polarized incident light. C1s XPS spectra have been recorded at a photon energy of 332 eV with an overall energy resolution of around 105 meV, and they have been calibrated with respect to the C1s binding energy of $CO_2$ (297.6 eV)[18].

NEXAFS spectra at the C K-edge were acquired by measuring the total ion yield (TIY) with an electron multiplier placed in the experimental chamber in front of the ionization region. The photon flux was measured simultaneously using a calibrated Si photodiode (AXUV100 IRD™) for the spectra normalization. The energy scale of the spectra was calibrated by taking simultaneous spectra of the samples and of $CO_2$, with the characteristic transition at 290.77 eV (C $1s \rightarrow \pi^*$, $CO_2$).[19,20] The photon energy resolution was around 65 meV.

The commercially available Sigma-Aldrich thiophene (purity ≥ 99%), 2,2'-bithiophene (97%) and 2,2':5',2–terthiophene (99%) have been used for analysis. 1T and 2T are liquid at room temperature and after several freeze-pump-thaw cycles of purification they were inserted in vacuum and dosed via a stainless steel variable leak valve. 3T sample is a solid and it was sublimated in vacuum using a custom built resistively heated furnace set at T=66° degrees after a purification treatment at 30°C for 12 hours.

The C1s XPS peaks were fitted using IGOR PRO™ and with an in-house written curve fitting routine. In order to reproduce the experimental lineshape, we have fitted the experimental data using a main peak followed by a progression of smaller peaks at higher binding energies. In this way we have built a function that accounts for an "adiabatic" peak accompanied by a vibrational "tail", and the function has been repeated for each different C sites ($C_i$). The lineshape for all $C_i$'s of one molecule was kept fixed, i.e. the widths of all sub-peaks were the same and their relative intensities were also fixed. NEXAFS spectra were analyzed using an analysis of the second derivative to determine the position of the spectral features.

**Computational details**

The geometrical structures of thiophene, 2,2'-bithiophene and 2,2':5',2"-terthiophene have been optimized at the density functional theory (DFT) level whit the LDA VWN functional[21] and the Triple Zeta Polarized (TZP) basis set of Slater type orbitals (STO) by the ADF package[22,23]. 2T and 3T have been considered in the *all-trans* conformations with a planar structure (with $C_{2h}$ and $C_{2v}$ symmetry in the GS, respectively). This choice is supported by a conformational analysis (performed with the TZP basis set and the PW86xPerdew functional) which indicates that the energy of the trans conformer is lower than cis by 0.79 kcal/mol for 2T and by 1.31 kcal/mol for 3T. The estimated conformer fractional populations have been then derived assuming a Boltzmann population distributions giving the following results : 0.80 (trans) and 0.20 (cis) for 2T at room Temperature (298 K); 0.88 (trans) and 0.12 (cis) for 3T at T= 339 K. Preliminary test calculations (not reported) on the C1s NEXAFS of the 2T cis conformer did not show any appreciable difference in the spectrum with respect to the trans conformer, so only trans conformers will be considered in the following.

The C1s NEXAFS and XPS spectra have been calculated at DFT level with the GGA PW86xPerdew functional [24] by the ADF program. For the NEXAFS spectra simulation, the core hole at each non-equivalent carbon center is modeled by the half core hole (HCH) also referred as the Transition Potential (TP) approximation [25]. In the TP computational technique half an electron is removed from the 1s orbital of the excited C atom, relaxing all the orbitals until self-consistency is obtained. This scheme includes most of the relaxation effects following the core hole formation and provides a single set of orthogonal orbitals useful for the calculation of the transition moments. The basis functions employed in the DFT-TP calculations consist of a very extended STO set for the core-excited carbon atom, in particular an even tempered Quadruple Zeta with three polarization and three diffuse functions (designed as ET-QZ3P-3DIFFUSE set in the ADF database), while the TZP basis set has been employed for the remaining atoms. Symmetry is properly reduced allowing core-hole localization.

The raw spectra are so calculated: the excitation energies are obtained as the differences between the eigenvalue of the virtual orbital and that of the 1s orbital calculated with TP configuration:

$$(1) \quad \Delta\varepsilon_{1s \to a} = \varepsilon_a - \varepsilon_{1s}$$

The excitation intensity is calculated with the oscillator strength that, within the dipole approximation, for excitation from the ground state (GS) $|\Psi_g\rangle$ to excited state $|\Psi_e\rangle$, is given by (atomic units used),

$$(2) \quad f_{g \to e} = \frac{2\Delta E_{g \to e}}{3} |M_{g \to e}|^2$$

$$(3) \quad M_{g \to e} = \langle \Psi_g | \hat{\mu} | \Psi_e \rangle$$

where $M_{g \to e}$ is the transition dipole moment with $\hat{\mu} = \sum_i r_i$ being the dipole operator, and $\Delta E_{g \to e} = E_e - E_g$ is the total energy difference. Taking into account the final-state rule [26,27] and the sudden approximation [11], Eq2 can be formulated at one-electron level, so that the oscillator strength is evaluated in terms of two molecular orbitals (MOs) $\psi_{1s}$(core) and $\psi_a$ of the final state, obtained with the DFT-TP scheme:

$$(4) \quad f_{g \to e} = \frac{2 \Delta \varepsilon_{1s \to a}}{3} |\langle \psi_{1s}(1)|r_1|\psi_a(1)\rangle|^2$$

The TP approach leads to a less attractive potential and the absolute transition energies are generally too large. In order to correct the NEXAFS energies, the raw spectra are calibrated by aligning the first transition energy $\Delta \varepsilon_{1s \to LUMO}$ (LUMO: lowest unoccupied MO) to that obtained from $\Delta$Kohn-Sham ($\Delta$KS) scheme [28,29], as difference between the total energy of the excited state ($E_{e_1}$) and the total energy of the ground state ($E_g$): $\Delta E_{g \to e_1} = E_{e_1} - E_g$. In order to get a pure singlet first core excited state, $E_{e_1}$ is calculated as:

$$(5) \quad E_{e_1} = 2E(|1s\alpha^1 \dots LUMO\beta^1|) - E(|1s\alpha^1 \dots LUMO\alpha^1|)$$

where $E(|1s\alpha^1 \dots LUMO\beta^1|)$ and $E(|1s\alpha^1 \dots LUMO\alpha^1|)$ are the total energies of two spin-polarized single-determinants with unpaired electrons in the 1s and LUMO orbitals (antiparallel and parallel, respectively). The excitation spectrum at each non-equivalent Carbon site of the molecule is obtained as a single calculation, then it is weighted by relative abundance and finally the total C1s NEXAFS spectrum is obtained by summing up the different contributions. In order to facilitate the comparison with the experiment, the raw spectra have been broadened by using a Gaussian lineshape with Full-Width-at-Half-Maximum (FWHM) of 0.4eV for 1T and 3T and of 0.3eV for 2T.

For the XPS simulation the core Ionization Potentials (IP) are determined at $\Delta$KS level in the following way:

$$(6) \quad IP_{1s} = E(|1s\alpha^1 \dots|) - E_g$$

where $E(|1s\alpha^1 \dots|)$ represents the total energy of a spin-polarized Full Core-Hole(FCH) state.

## Results and discussion

### C 1s XPS

Photoemission spectra of the C1s core levels are shown in Fig.2 together with the analysis of the experimental lineshapes performed by the fitting procedure described in the experimental section. The C1s photoemission spectra of thiophenes are known to be affected by a pronounced vibrational envelope [30]. Remarkably, while the vibrational tail is very intense for 1T, it decreases remarkably for 2T and is very low for 3T, though it is still necessary for an acceptable fitting. The intensity of the vibrational tails is about 40% for 1T, 27% for 2T and only 8% for 3T.

For the assignment of the spectral features we have compared them to the calculated ΔKS binding energies (BE), which are displayed as vertical bars in the Figure 2 and listed in the Table 1 together with the results of the fit.

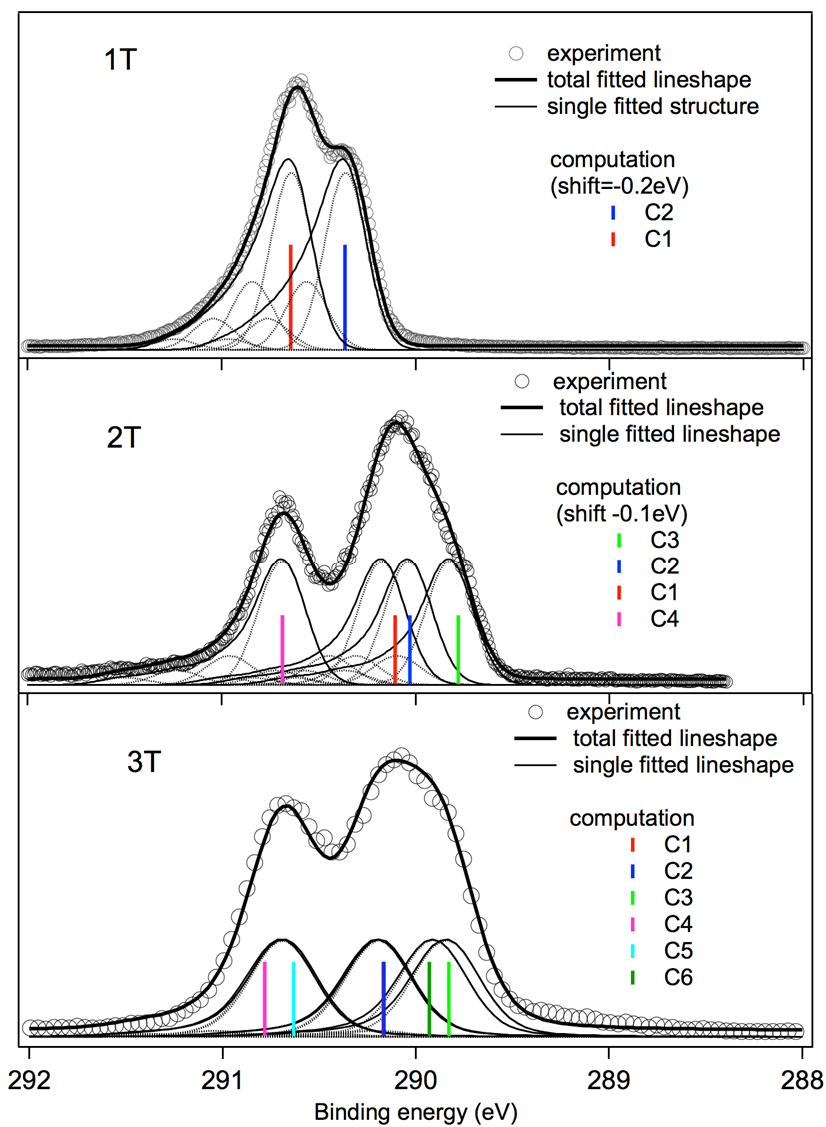

*FIG. 2) C 1s XPS spectra of 1T (upper panel), 2T (central panel) and 3T (lower panel): experimental data (circles) are shown together with the results of the peak fitting procedure described in the text. The "adiabatic" peaks $C_i$ accompanied by their vibrational progression are represented by the thin solid lines, while the solid thick lines are the total sum of the fitted lineshape. The vertical colored lines are the BE calculated at the ΔKS level; the line intensity is proportional to the multiplicity of the C atoms and the energy scale for 1T and 2T has been shifted*

*respectively by -0.2 and -0.1 eV. The calculated C1 and C2 BE of 3T are almost degenerate and overlap in the plot.*

It would be too difficult to calculate the absolute intensity of the XPS lines since the unbound photoelectron wave function should be employed in the dipole transition moment. Therefore, we simply assume that the intensity of each C1s line is proportional to the number of equivalent carbon atoms (in each of the present molecules all the C atoms have the same site multiplicity of 2). Following the calculated BE, it is possible to define two different groups of carbon atoms: the first one at lower binding energy corresponds to the carbon atoms not bonded to a sulfur atom, while the second group at higher energies identifies the carbons adjacent to the sulfur. We can also note that in 2T and 3T the highest BE are relative to carbons bound to S atom and connecting two rings.

Table 1. Calculated BE with corresponding carbon site (first column).
Experimental BE extracted with the fitting procedure described in the text (last column)

| | BE (eV) | |
|---|---|---|
| **Thiophene** | ΔKS | EXP fit |
| C1 | 290.862 | 290.60 |
| C2 | 290.583 | 290.35 |
| | | |
| **Bithiophene** | ΔKS | EXP fit |
| C4 | 290.987 | 290.70 |
| C1 | 290.404 | 290.19 |
| C2 | 290.330 | 290.07 |
| C3 | 289.078 | 289.89 |
| | | |
| **Terthiophene** | ΔKS | EXP fit |
| C4 | 290.781 | 290.70 |
| C5 | 290.633 | 290.69 |
| C1 | 290.167 | 290.20 |
| C2 | 290.166 | 290.19 |
| C6 | 289.929 | 289.91 |
| C3 | 289.828 | 289.84 |

The higher BE of carbons bound to the sulfur is consistent with the higher electronegativity of sulfur compared to carbon, as well as to the charge transfer from sulfur to C atom not bounded with it through resonance in the π system. The further BE increase for the C atom connecting the rings reproduces a trend already observed in polycyclic aromatic hydrocarbons for which the calculations provides higher BE for C atoms non bound to any hydrogen atom compared to external carbons bound to hydrogen atoms[31].

Starting from the 1T spectrum, the measurements can be obviously interpreted as derived from the two chemically shifted carbon lines: the part at higher energy assigned to C1, atom close to sulfur, separated by 279 meV from the C2 subspectrum. This chemical shift is in good agreement with the experimental value of 0.25 eV. The intensity distribution between the two observed peaks does not reproduce the theoretical statistical ratio of the carbon atoms (1:1) because it is affected by strong vibrational effects: the C1 peak shows a higher intensity as compared to C2 because it overlaps with the vibrational tail of the latter, as shown in Fig.2. The most pronounced vibrational components, in particular deriving from the S-C1 bond stretching, are present in the C1 1s spectrum compared to C2, as analyzed by Giertz et al. [30].

For 2T, the calculations assign the higher energy peak to the C1s α-carbon atom (C4), connecting the two thiophene rings; the energy shift with respect to the other C atom (C1), bound to sulfur and also to an hydrogen atom, is quite significant (583 meV) so the calculated C1 line contributes to the lower energy peak together with the lower energy lines relative to the C atoms not bonded to sulfur (C2 and C3). The lower binding energy of C3 compared to C2 can be ascribed to an addition of valence electron charge on the C3 site as a result of changes in bonding for the aromatic conjugation induced by the second ring. The theoretical statistical ratio of these two groups of C atoms is equal to 1:3 and qualitatively accounts for the relative height of the two experimental peaks. We note also the significant decrease of the energy shift between C1 and C2, 74 meV to be compared with 279 meV of 1T. A similar trend is found also for the calculated BE of terthiophene: the higher BE refer to the two C atoms (C4, C5) which connect the rings and lose their equivalence in this molecule, while the lowest BE are found for the C atoms not bonded to S atom (C2, C3 and C6 respectively). The energy separation between C1 and C2 reduces to such an extent that the two BEs are almost degenerate while the energy shift between the (C4, C5) group and the (C1, C2) one (540 meV) is similar to that found in 2T as well as the lowest BE calculated for the C3 and C6 sites. The presence of a second and third rings introduces slight different aromaticity effects so that the C1 and C3 sites are destabilized by an increase of charge density, in line with the calculated decrease of C1 and C2 binding energies along the series. The first peak is, therefore, assigned to the binding energies of C4, C5 sites while all the other carbon lines contribute to the lowest energy experimental peak. The statistical ratio is therefore 1:2 which matches the relative areas of the experimental structures.

## C 1s NEXAFS

The results of the computed excitation energies and oscillator strengths in C 1s NEXAFS spectra are collected in Table 2 and reported in Figure 3 where the calculated profile of the total spectrum is shown in black, while the colored lines refer to the transitions relative to the Carbon non-equivalent sites. The site-resolved excitations facilitate the analysis of the transitions and the attribution of the spectral features to specific portions of the molecule. The theoretical C1s ionization thresholds are also shown in the figures in order to distinguish the below-edge region of the spectrum, where the present discrete orbital description is adequate, from the above-edge region, where such an approach determines a discretization of the non-resonant continuum that is in part an artifact of the calculation, so that only qualitative information could be extracted above the ionization threshold. Figure 3 also reports the experimental profiles useful for a comparison with the calculated ones. A detailed comparison between theory and experiments is discussed later in connection to Fig. 4.

We first explain, in detail, the nature of the spectral features. The calculated total C1s spectrum of the 1T molecule (Table 2 and upper panel of Figure 3) is characterized by a first sharp peak (A, at 285.65 eV) which is contributed by the two C1s → LUMO transitions from both C1 and C2 carbon sites. The LUMO orbital ($1\pi^*$) is appreciably localized at the Sulphur atom, because of the considerable aromatic character of 1T, and this is reflected in the lower intensity of the C2 1s→$1\pi^*$ compared to the C1 1s→$1\pi^*$ transition. The two calculated $\pi^*$ transitions are substantially degenerate and this result could appear in disagreement with the energy split of 0.28 eV found for the C1 and C2 XPS binding energies (see Fig.1 and Table 1) as well as with the broader shape with a shoulder to the high energy side of the experimental peak, which could suggest the contribution of two not degenerate electronic transitions. To investigate the underlying reasons of this disagreement with the experiment, we have performed a simulations of the C1s spectrum of the 1T employing also another exchange-correlation functional, in particular the B3LYP [32] one. No significant variation in the energy separation between the first two $\pi^*$ transitions is obtained as well as in the overall C1s NEXAFS spectrum, suggesting that the description of the electronic structure is correct. The calculated degeneracy of the two $1\pi^*$ transitions can be explained as a final state relaxation effect of the LUMO virtual orbitals: when the half-hole is created on C1 the relaxation of the LUMO is stronger compared to that of the LUMO in the analogous calculation on the C2 site. This differential relaxation compensates for the initial state energy shift (initial state effect) and might be related to a greater aromatic stabilization of the C1 LUMO compared to C2 LUMO. A possible cause that might explain the broader peak of the experiment is the presence of vibronic effects, which have found to influence the XPS C1s profile (see Ref. 30 and our discussion of the XPS lineshape), and are not included in the present computational model. We will comment on this aspect at the end of this section, where the experimental and theoretical data are compared in Fig.4.

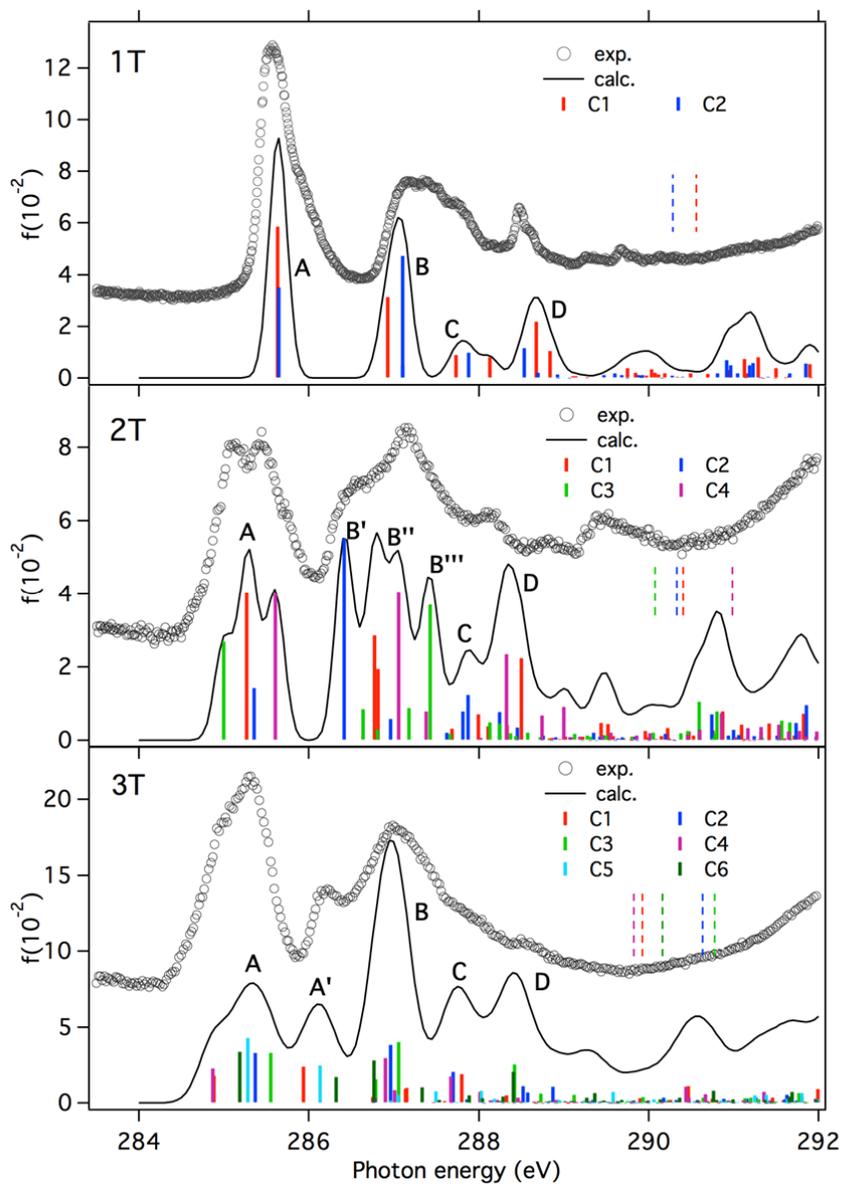

*Figure 3. C 1s NEXAFS spectra of 1T (upper panel), 2T (central panel) and 3T (lower panel): calculated lineshape (black solid line) with partial $C_i$ contributions (thick colored vertical bars). The ΔKS C1s ionization thresholds are also shown (colored vertical dashed bars). The energy scale of the experimental results for 1T has been shifted of +0.35 eV in order to match the first calculated peak.*

The second calculated peak B is a superposition of two excitations: from C1 (at 286.93 eV) towards a σ* (S-C) virtual orbital, which is largely contributed by the S lone pairs with a smaller $2p_x$ in plane (x) component of C1, and from C2 into the second π* (2π*) orbital at 287.11 eV, which is localized on the four carbon atoms. The analogous 1s→2π* transition from C1 is weaker and lies at 287.7 eV, therefore, above the corresponding transition from C2, and contributes to the small peak C. The lower intensity of C1 1s→2π* compared to C2 1s→2π* transition is due to the higher localization of the 2π*final orbital on the C atoms not directly bonded to the Sulphur. The other two weak transitions contributing to peak C derive from two equivalent C1 and C2 transitions into diffuse orbitals. The last peak before threshold (peak D at 288.7 eV) corresponds to transitions to final C3p Rydberg MOs with some σ*(C-H) contribution; the mixed valence-Rydberg nature of these transitions is consistent with their significant intensity.

Table 2. Calculated excitation energies (eV) and oscillator strengths of 1T.
Only the main transitions are reported.
Experimental energies extracted with the fitting procedure described in the
text are reported in the last column. The transitions marked with (*)
in the experimental data indicate a vibrational feature

| Peak | Site | E (eV) | fx10$^2$ | Assignment | EXP |
|---|---|---|---|---|---|
| A | C1 | 285.64 | 5.86 | LUMO (1π*) | 285.15 |
|   | C2 | 285.65 | 3.50 |  | 285.30 |
|   |    |        |      |  | 285.58 (*) |
|   |    |        |      |  | 285.75 (*) |
| B | C1 | 286.93 | 3.12 | σ*(S-C) | 286.68 |
|   | C2 | 287.11 | 4.73 | 2π*(C=C) | 286.79 |
|   |    |        |      |  | 287.04 (*) |
|   |    |        |      |  | 287.17 (*) |
| C | C1 | 287.73 | 0.89 | 2π*(C=C) | 287.40 |
|   | C2 | 287.88 | 0.97 |  | 287.52 |
|   | C1 | 288.13 | 0.79 | Rydberg | 287.81 |
| D | C2 | 288.53 | 1.15 |  | 288.13 |
|   | C1 | 288.68 | 2.17 | Rydberg+σ*(CH) | 288.25 |
|   | C1 | 288.84 | 1.05 |  | 288.30 |

The assignment of the features substantially agrees with that previously published [33,34].

The 2T spectrum (Table 3 and middle panel of Figure 3) shows an increased complexity due to the presence of four non-equivalent carbon sites as well as of the increased number of low-lying virtual π* orbitals. The first structure (A, around 285.5 eV) is assigned to transitions from the four C1s non-equivalent carbon atoms into the LUMO orbital, which is contributed by the valence C 2p$_z$ and S 3p$_z$ components and, therefore, comparable in nature to the LUMO orbital of 1T. These excitations cover an energy range of about 0.6 eV due to the splitting of the C1s binding energies and their convolution gives rise to a double-peak shape with a shoulder. These excitation energies reflect only partly the relative BE of the core orbitals; in particular, each C1s α-carbon atom (C4), connecting the two thiophene rings, feels the proximity of the S atom and of two C atoms with a stronger depletion of electron density than the other C atoms of the rings. The C4-1s orbitals are therefore less screened and a larger excitation energy is predicted for them which contribute to the higher energy side of peak A, while the lowest energy of the C3-1s LUMO excitation accounts for the most screened C1s orbital compared to the other carbon atoms, in agreement with the calculated binding energies (see Fig.1). The localization of the LUMO mainly on C1 and C4 is responsible for the stronger intensity of the relative transitions to the LUMO compared to the other two C1s excitations of the A peak. The second calculated feature (B), around 287 eV, displays a three peaked shape (denoted as B', B'' and B''') as a result of the convolution of the many intense transitions falling in this energy range. These transitions roughly correspond to those contributing to the 1T B peak, namely towards the $2\pi^*$ and the $\sigma^*$ (S-C) antibonding orbitals. The $2\pi^*$ orbital resembles the LUMO composition maintaining a significant S $3p_z$ component consistent with an increased aromatic character compared to 1T. In particular, the B' component of the peak B derives from the C2-1s→$2\pi^*$ transition (at 286.42 eV) while analogous transitions from the C1 and C3 sites (at 286.77 eV and 287.43 eV) contribute to the B'' and B''' components respectively. The strongest $\pi^*$ transitions still involve the C atoms not directly bonded to a Sulphur atom, as in 1T. The excitations to the $\sigma^*$ (S-C) antibonding orbitals contribute significantly to B'' component and derive from the C1 and C4 carbon sites directly bonded to a sulphur atom (at 286.77 eV and 287.06 eV respectively). The assignment of the main features substantially agrees with multilayer films of 1T and 2T deposited on Ag(111)[34] and for a monolayer of 3T deposited on Ag(110)[35]. Peak C and D should be characterized as superposition of valence and Rydberg excitations. The most intense are the valence transitions towards final orbitals of $\pi^*$ character (from C2 site at 287.88 eV, peak C and from C4 site at 288.33 eV, peak D); their reduced intensity compared to the lowest energy $\pi^*$ transitions reflects the decrease of the valence $C2p_z$ contribution of the C1s excited site in the higher $\pi^*$ virtual MO. The less intense lines are assigned to transitions into diffuse orbitals of mainly C3p-Rydberg character.

Table 3. Calculated excitation energies (eV) and oscillator strengths of 2T.
Only the main transitions are reported.
Experimental energies extracted with the fitting procedure described in the
text are reported in the last column.

| Peak | Site | E (eV) | f x10$^2$ | Assignment | EXP |
|---|---|---|---|---|---|
|   | C3 | 285.00 | 2.68 |   | 285.02 |
| A | C1 | 285.27 | 4.03 |   | 285.12 |
|   | C2 | 285.36 | 1.42 | LUMO (1π*) | 285.47 |
|   | C4 | 285.61 | 4.00 |   | 285.78 |
| B' | C2 | 286.42 | 5.52 | π* | 286.35 |
|   | C1 | 286.78 | 2.86 | σ*(S-C) | 286.53 |
| B'' | C1 | 286.82 | 1.94 | π* | 286.84 |
|   | C4 | 287.06 | 4.05 | σ*(S-C) | 287.14 |
| B''' | C3 | 287.43 | 3.69 | π* | 287.52 |
|   | C2 | 287.81 | 0.80 | Rydberg |   |
| C | C2 | 287.88 | 1.24 | π* | 287.84 |
|   | C1 | 287.99 | 0.71 | Rydberg |   |
| D | C4 | 288.33 | 2.34 | π* |   |
|   | C1 | 288.50 | 2.23 | Rydberg + σ*(CH) | 288.17 |

The complexity further increases in the 3T spectrum (lowest panel of Fig.3 and Table 4); in this molecule there are six nonequivalent C atoms and an even greater number of low-lying π* orbitals. The peak A arises from the C1s→LUMO transitions from all the six carbon sites whose contributions are highlighted in Figure 2. The larger excitation energies are predicted for the C4 and C5 carbon sites which are less screened being directly bounded to the sulfur atoms and connecting two rings, as already found for the 2T molecule, and in agreement with the energy position of the calculated IPs. However, the differential relaxation effects on the π* low-lying orbitals, depending on the localization of the carbon core hole, prevent an analysis of the following features based on a regular energy shift of the site-resolved excitation spectra following the energy position of the IPs.

The feature A' corresponds to transitions to the second $\pi^*$ orbital ($2\pi^*$) from C6, C2 and C1 sites while the transition from C4 site is shifted at higher energy (286.79 eV, peak B). We do not observe analogous transitions from C3 and C5 sites because the $2\pi^*$ relative final orbitals have a negligible C2$p_z$ contribution of the carbon excited site. The B peak arises from several transitions to $\pi^*$ orbitals overlapped on the stronger $\sigma^*$ (S-C) transitions from the C1, C4 and C5 sites; this attribution confirms the mixed nature of peak B found also for 1T and 2T. The C and D peaks are still contributed by transitions to virtual orbitals with $\pi^*$ character; the calculations indicate that these higher energy $\pi^*$ orbitals are mostly localized on carbon atoms with a reduction of the S3$p_z$ contribution as well as of the conjugation among the rings. The progressive intensity decrease of these transitions reflects the general reduction of the C2$p_z$ valence character of the final MOs; in the region of peak D also transitions to diffuse MOs with Rydberg C3p components are present with lower intensity.

Table 4. Calculated excitation energies (eV) and oscillator strengths of 3T. Only the main transitions are reported.
Experimental energies extracted with the fitting procedure described in the text are reported in the last column.

| Peak | Site | E (eV) | f x10$^2$ | Assignment | EXP |
|---|---|---|---|---|---|
|  | C3 | 284.87 | 2.26 |  |  |
|  | C6 | 284.88 | 1.77 |  | 284.87 |
|  | C1 | 285.18 | 3.38 |  | 285.00 |
| A | C2 | 285.28 | 0.92 | LUMO (1π*) | 285.35 |
|  | C5 | 285.37 | 3.27 |  | 285.42 |
|  | C4 | 285.55 | 3.29 |  |  |
|  | C6 | 285.94 | 2.39 |  |  |
| A' | C2 | 286.13 | 4.27 | 2π* | 286.15 |
|  | C1 | 286.32 | 1.70 |  |  |
|  | C1 | 286.77 | 2.80 | σ*(S-C) |  |
|  | C4 | 286.79 | 1.56 | 2π* |  |
|  | C3 | 286.90 | 2.96 | π* |  |
|  | C5 | 286.96 | 3.81 | σ*(S-C) | 286.62 |
| B | C2 | 287.05 | 2.44. | π* | 286.91 |
|  | C4 | 287.06 | 3.99 | σ*(S-C) | 287.32 |
|  | C6 | 287.15 | 0.97 | π* |  |
|  | C1 | 287.34 | 1.01 | π* |  |
|  | C3 | 287.67 | 1.75 | π* | 287.90 |
| C | C5 | 287.70 | 2.05 | π* | 288.04 |
|  | C6 | 287.80 | 1.88 | π* |  |
|  | C1 | 288.41 | 2.03 | Rydberg |  |
|  | C4 | 288.42 | 2.52 | π* | 287.55 |
| D | C5 | 288.53 | 1.11 | π* | 288.66 |

|     | C5 | 288.87 | 1.07 | π* |

We finally address the evolution of the spectral features of the oligothiophenes when increasing the number of thiophene rings. The lower energy features (A and B) maintain their nature along the series, despite the greater complexity and the enlargement of the peaks. The first π* peak (C1s transitions to the LUMO) shifts to lower energy (about 0.5 eV) from 1T to 2T while it does not change further from 2T to 3T: this trend can explain a stabilizing effect due to the aromaticity, which is stronger in passing from one to two rings than in adding a third ring. The σ* (S-C) transitions maintain their energy almost constant along the series because the involvement of a single bond is not influenced by aromaticity and always falls in the region of peak B together with higher π* transitions. The number of overlapping transitions increases at higher energies (peaks C and D) preventing a strict correspondence along the series.

In Figure 4 the C 1s NEXAFS spectra computed by the TP-DFT scheme are compared with the gas phase experiments and with the analysis of the experimental lineshapes using the second derivative. The 1T calculated profile (upper panel) has been shifted on the energy scale (+0.35eV) to match the first peak of the experiment. In this way, the relative energy shift among the calculated transitions, which actually represents the most significant observables, is preserved.

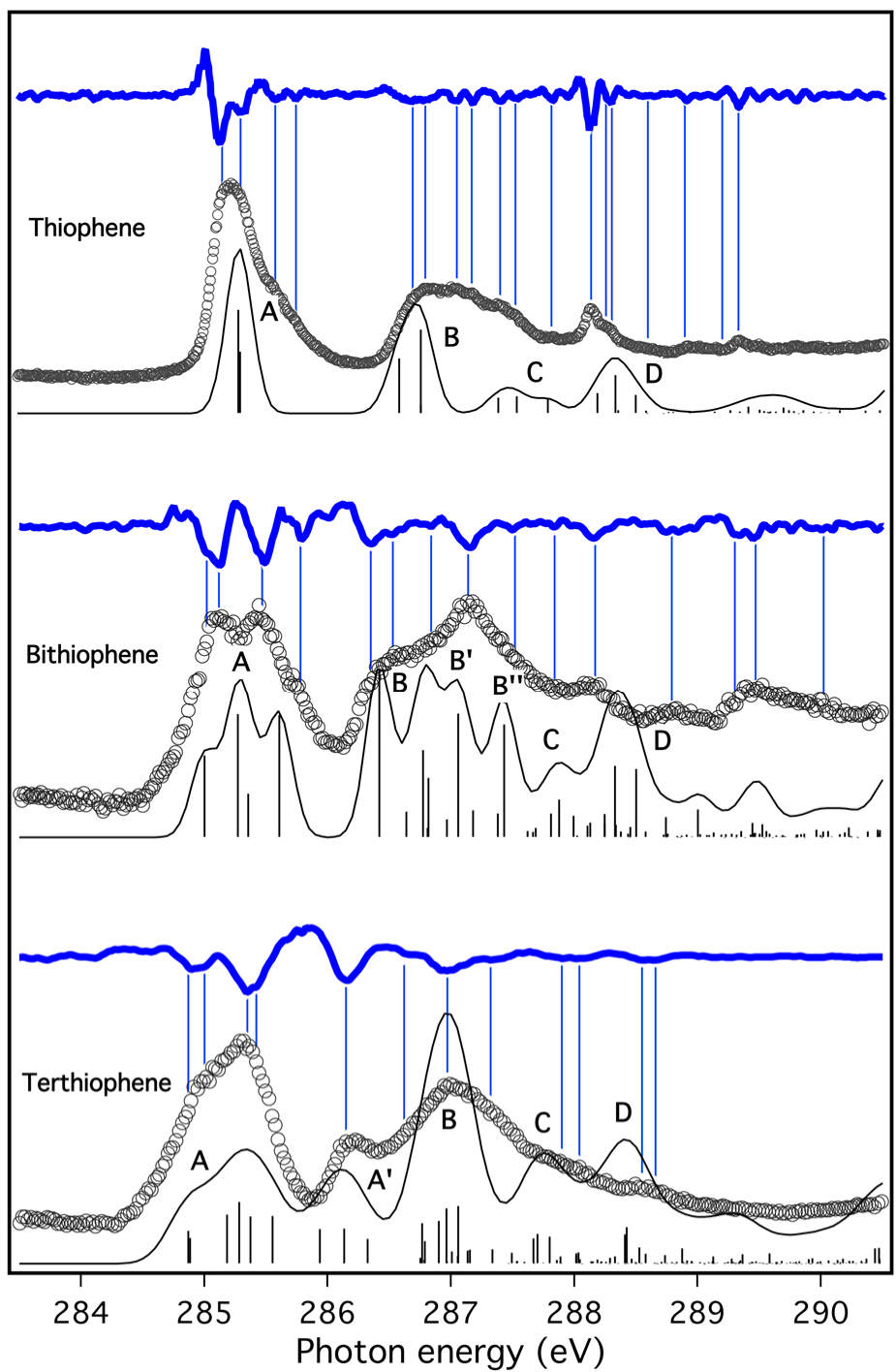

*Figure 4. C 1s NEXAFS spectra of 1T (upper panel), 2T (central panel) and 3T (lower panel): experimental data (circles) and calculated DFT-TP results (black solid lines and vertical bars). The energy scale of the DFT-TP results for 1T has been shifted of -0.35 eV in order to match the first experimental peak. At the top of each spectrum the second derivative of the experiment, for the determination of peak positions, is also shown (blue solid line); the main minima are indicated with the thin vertical blue bars.*

A general good agreement is reached between experiment and theory, in particular the theoretical results correctly describe the main features of the experimental curves and the energy separation among the peaks. A clear correspondence between calculated and experimental peaks is apparent in the 1T spectrum; the major discrepancy concerns the broader shape of the experimental peak A of 1T with respect to the theoretical one, as already highlighted. We have hypothesized that this disagreement can be caused by vibronic effects responsible for the asymmetry with an high energy tail of the experimental peak. This hypothesis can be supported by the results of our recent study on C1s NEXAFS spectra of pyridine and its fluorinated derivatives [36], which demonstrated that vibronic effects are responsible for strong intensity distributions among the C1s→π* excitations of the three nonequivalent C atoms and their inclusion in the calculations are mandatory for a quantitative description of the experimental lineshape of the first double-peak band. In particular, the C1s→π* calculated vibronic progressions of each nonequivalent C atom show an intense 0-0 transition followed by less intense vibronic transitions (shifted by about 0.2 eV) which give rise to a tail extending up to about 1 eV. The overlap of the vibronic progressions is responsible for the broadening observed in particular at the higher energy side of the double-peak. Although a generalization of the pyridine behavior to another class of small heterocyclic molecules like a five member thiophene ring is not straightforward, we think that the correct reproduction of the first peak lineshape in thiophene can be more likely caused by the neglect of vibrational effects in the computational simulation than to a deficiency in the electronic structure description. This issue deserves a future work to be explored and assessed with more confidence. In the present hypothesis, it is straightforward to attribute the first two experimental energy values obtained by the fitting procedure (285.15 and 285.30 eV, Table 2) to the C1 and C2 1s transitions to 1π* orbitals, and the third at 285.58(*) eV and fourth at 285.75(*) eV to their vibrational progression.

Similarly for group A, the features in group B at 286.68 eV and 286.79 eV correspond to adiabatic peaks while 287.04 (*) eV e 287.17(*) eV are vibrational. The following peaks at 287.40 eV, 287.52 eV and 287.81 eV correspond to the computed ones and most probably they overlap with the vibrational structures. The ones at 287.73 eV, 287.88 eV and 288.13 eV correspond to calculated Rydberg peaks. Vibronic effects could also be responsible for other discrepancies between theory and experiment in the 2T and 3T spectra, in particular as concerns the first peak. The intensity distribution of the experimental double-peak feature of the bithiophene spectrum is not properly reproduced by the calculation while in the 3T spectrum the theory underestimates the first peak intensity with respect to the second peak. We would like to underline that the presence of other minor conformers of 2T[37] and 3T [38] is also possible, as discussed in the computational section, which can make more difficult a ono-to-one match of the theoretical and experimental spectral features. In summary, a comparison between the TP-DFT electronic calculations and experiment is fully satisfactory as far as the relative excitation energies are concerned, while the intensity distribution is less quantitative. We tend to ascribe this problem to the neglect of vibrational effects in the computational approach, also on the basis of our previous vibrationally resolved studies on NEXAFS C1s spectra of both simple aromatic molecules[36] and polycyclic aromatic hydrocarbons[31].

**Conclusions**

The electronic structure of thiophene, 2,2'-bithiophene and 2,2':5',2''-terthiophene have been thoroughly investigated by means of NEXAFS and XPS spectroscopy at the C K-edge. The rationalization of the experimental results has been successfully guided by the outcomes of calculations performed in the framework of DFT. It has been shown that the TP-DFT calculations are suitable to assign unambiguously the NEXAFS experimental features of systems of increasing complexity where strong electronic effects related to the aromatic stabilization, as well as inductive and relaxation effects generated by the presence of the S atom, are present. This demonstrates the ability of the TP-DFT scheme to describe in a balanced way electronic effects of such a different nature (aromatic stabilization, inductive and relaxation effects). The trend in the lower energy experimental structures of the three molecules points out the stronger stabilizing effect due to the aromaticity in passing from one to two rings compared to the addition of the third ring in terthiophene. Some discrepancies between theory and experiment are observed in the energy region of the first $\pi^*$ peak where the vibrational effects could play a role in the intensity distribution to properly fit the experimental spectral shape.

In the XPS the computation of the binding energy allows to discriminate the different carbon sites both on the basis of the bound with the Sulphur atom as well as on the differential aromaticity effects in 2,2'-bithiophene and 2,2':5',2''-terthiophene which split the BE according to whether the C atoms are bound or not to hydrogen. For XPS the theoretical statistical ratio of the carbon atoms is not always able to account for the intensity distribution between the observed peaks. This disagreement has been ascribed to complex vibrational effects which is not included in the present computational scheme. The capability of a conjugated system to transduce electronic effects depends on the delocalization of the charge carriers along the molecular chain, and to the best of our knowledge, this is the first methodic survey on the electronic structure and charge dynamics experienced by the main constituents of oligothiophene systems. In spite of the great importance of these systems to build more extended systems, which are the building blocks for the formation of polythiophenes, the understanding of the structure-related properties is still lacking and require more effort in basic research.

## Supplementary Material

See supplementary material for the pictures of the LUMO orbital of thiophene, 2,2'-bithiophene and 2,2':5',2''-terthiophene molecules.

## Acknowledgments

This work has been supported by Università degli Studi di Trieste, Finanziamento di Ateneo per progetti di ricerca scientifica, FRA 2014.

A.G. would like to acknowledge Elettra-Sincrotrone Trieste for providing financial support as Italian Funded Users to attend shifts assigned to the proposals n. 20140325 and n. 20145350 at the Gas Phase beamline.

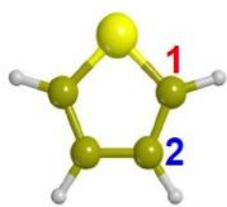 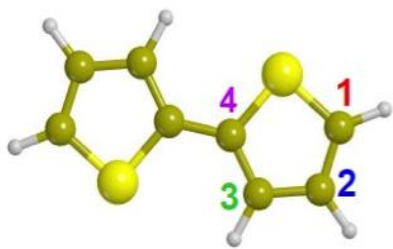 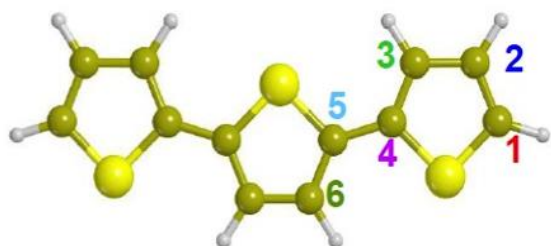

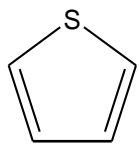 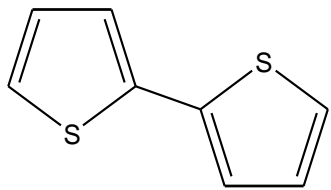 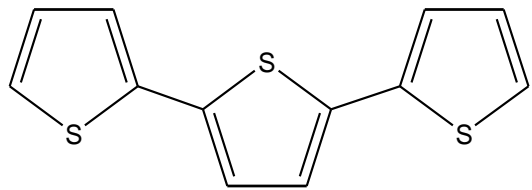

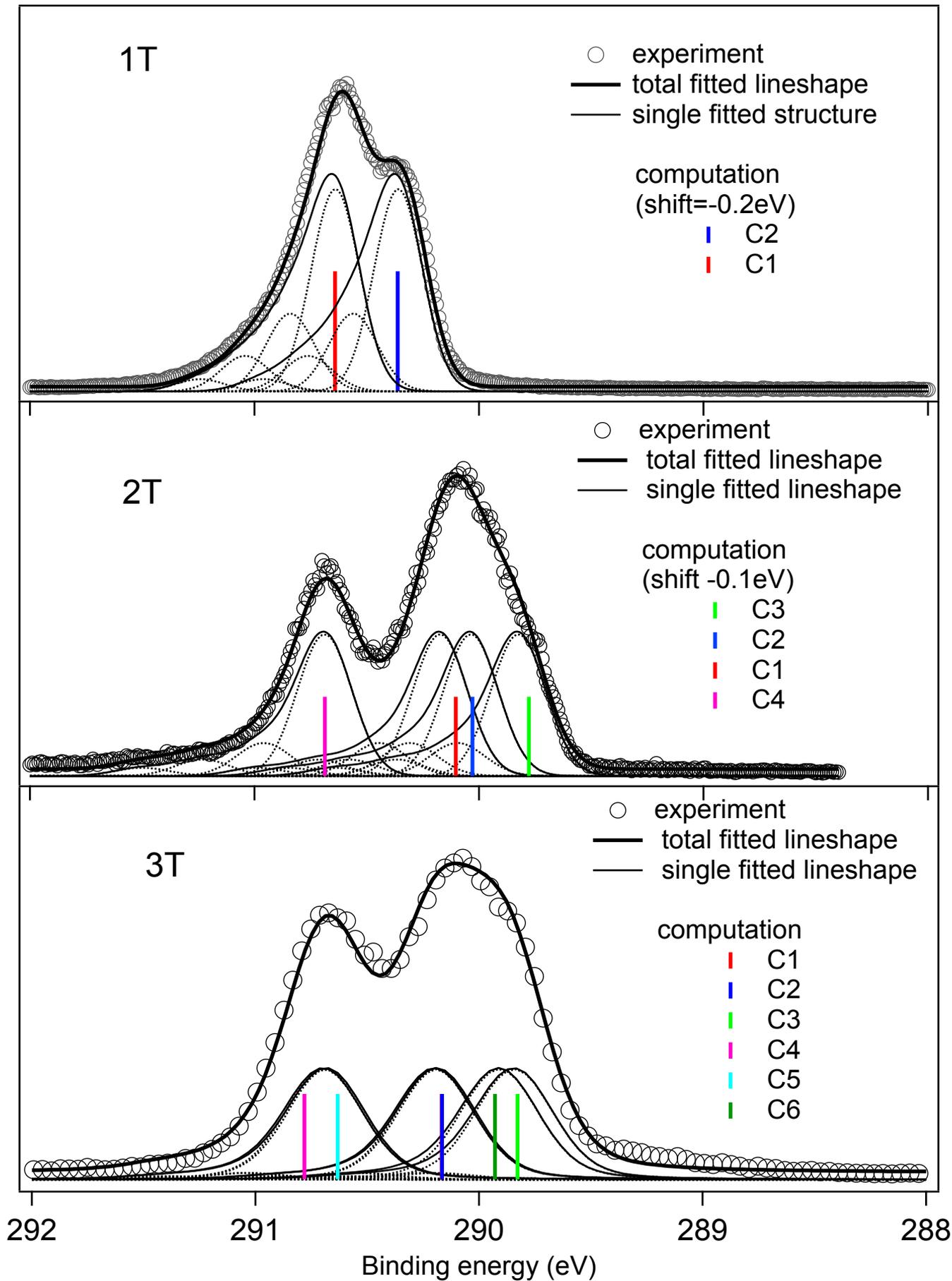

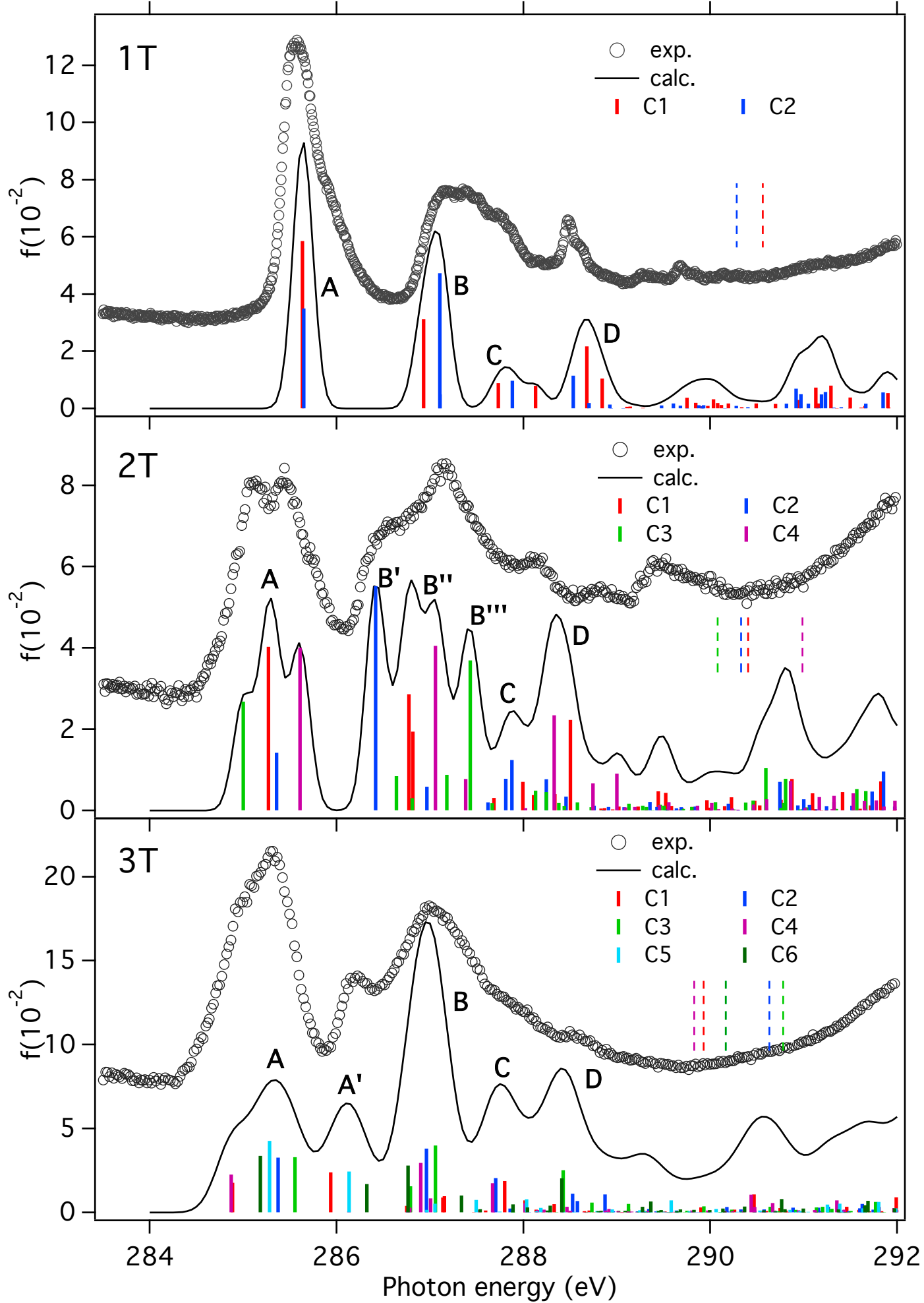

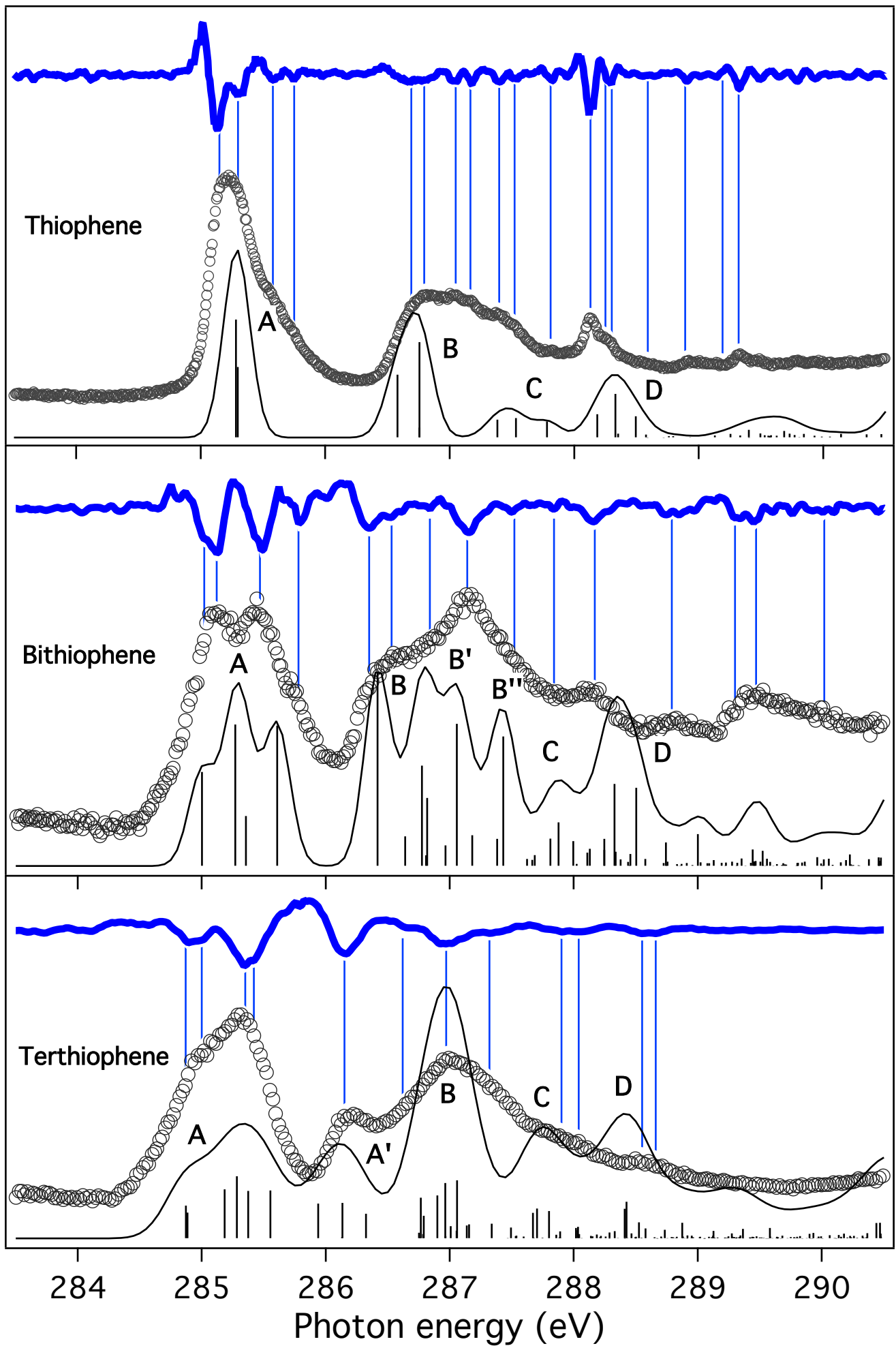

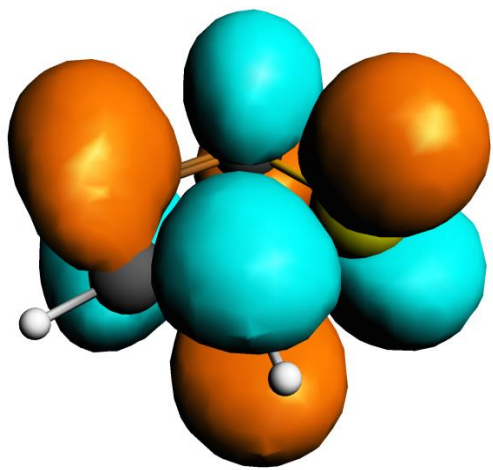
**1T**

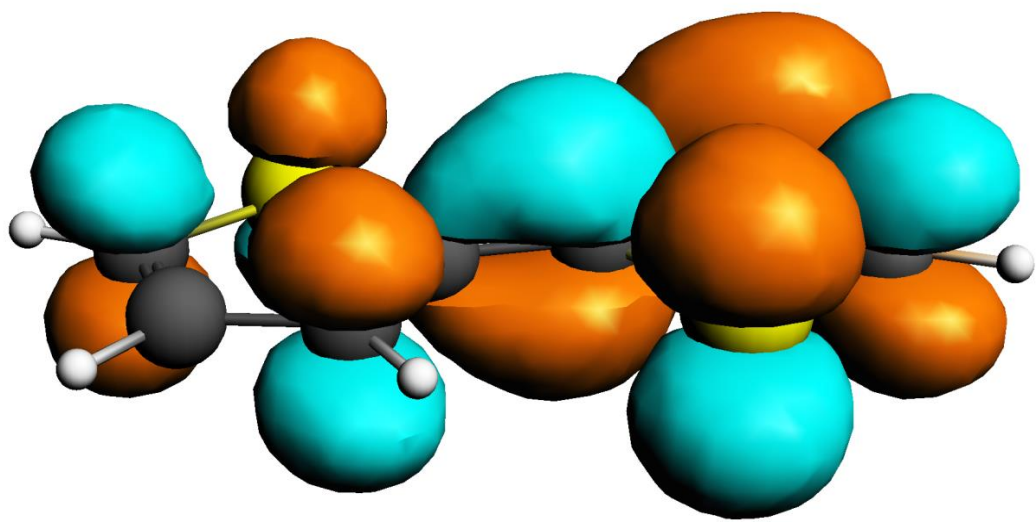
**2T**

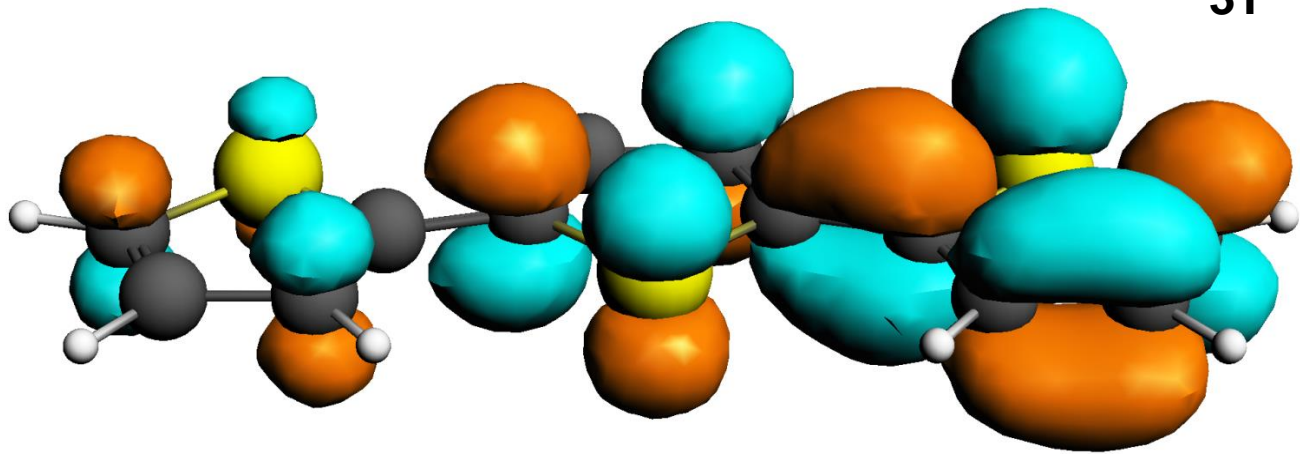
**3T**

3D plot of the LUMO orbital of 1T, 2T and 3T from the DFT-TP relaxed calculation on the C1 site.